\documentclass[iop,usenatbib]{emulateapj}

\def\1{\c{c}}
\def\2{\c{C}}
\def\3{\.{I}}
\def\4{\"{a}}
\def\5{{\i}}
\def\6{$\beta$}
\def\7{\"{o}}
\def\8{\"{O}}
\def\9{\c{s}}
\def\0{\c{S}}
\def\*{\"{u}}
\def\?{\"{U}}
\def\;{\u{g}}
\def\:{\u{G}}
\usepackage{color}

\slugcomment{The Astrophysical Journal Draft Version March 22, 2015}

\shorttitle{Searching for RP in the Gamma-ray Emitting SNR G349.7$+$0.2}
\shortauthors{Ergin et al.}

\begin{document}

\title{Searching for Overionized Plasma in the Gamma-ray Emitting Supernova Remnant G349.7$+$0.2}

\email{$^{\star}$ tulun.ergin@tubitak.gov.tr}
\thanks{ $^{\dagger}$ Now at Harvard-Smithsonian Center for Astrophysics, 60 Garden St., Cambridge, MA 02138, USA.}
\thanks{$^{\ddagger}$ Now at Tata Institute of Fundamental Research, Mumbai, 400 005, India}
\author{T. Ergin$^{\star}$, A. Sezer$^{\dagger}$}
\affil{TUBITAK Space Technologies Research Institute, ODTU Campus, 06531, Ankara, Turkey}
\affil{Bogazici University, Physics Department, Bebek, 34342, Istanbul, Turkey}
\author{L. Saha$^{\ddagger}$, P. Majumdar}
\affil{Saha Institute of Nuclear Physics, Kolkata, West Bengal 700064, India}
\author{F. G\7k}
\affil{Akdeniz University, Faculty of Education, Department of Secondary Science and Mathematics Education, Antalya, 07058, Turkey}
\author{E. N. Ercan}
\affil{Bogazici University, Physics Department, Bebek, 34342, Istanbul, Turkey}

\begin{abstract}
G349.7$+$0.2 is a supernova remnant (SNR) expanding in a dense medium of molecular clouds and interacting with clumps of molecular material emitting gamma rays. We analyzed the gamma-ray data of Large Area Telescope on board {\it Fermi Gamma Ray Space Telescope} and detected G349.7$+$0.2 in the energy range of 0.2$-$300 GeV with a significance of $\sim$13$\sigma$ showing no extended morphology. Modeling of the gamma-ray spectrum revealed that the GeV gamma-ray emission dominantly originates from the decay of neutral pions, where the protons follow a broken power-law distribution with a spectral break at $\sim$12 GeV. To search for features of radiative recombination continua in the eastern and western regions of the remnant, we analyzed the {\it Suzaku} data of G349.7$+$0.2 and found no evidence for overionized plasma. In this paper we discuss possible scenarios to explain the hadronic gamma-ray emission in G349.7$+$0.2 and the mixed morphology nature of this SNR.
\end{abstract}


\section{Introduction}
The Galactic supernova remnant (SNR) G349.7$+$0.2 has been long suggested to be on the edge of the Milky Way ($\sim$22 kpc), following the H\,{\sc i}  absorption measurements \citep{caswell1975} and the OH (1720 MHz) maser observations \citep{frail1996}. Recently, based on a better understanding of the Galactic Center's structure and using the 1420 MHz radio continuum and 21 cm H\,{\sc i} emission data sets from the Southern Galactic Plane Survey, \citet{tianleahy2014} estimated the distance of G349.7$+$0.2 to be $\sim$11.5 kpc.

In the radio waveband, the central spectral index value of $-$0.47 $\pm$ 0.06 suggests that G349.7$+$0.2 is a shell-type SNR. However, due to the enhancements towards the eastern and southern parts of its 2\farcm5 diameter shell, it was found to be an unusual shell-type SNR lacking a central cavity and a ring-like structure \citep{shaver1985}.

In X-rays, G349.7$+$0.2 was first detected in the Galactic plane survey of {\it ASCA} \citep{yamauchi1998}. The analysis of the {\it ASCA} data \citep{slane2002} showed that the spectrum was well fit by a single-temperature non-equilibrium ionization (NEI) model and they estimated the age of G349.7$+$0.2 to be $\sim$2800 yr. \citet{lazendic2005} analyzed the {\it Chandra} data of the whole SNR as well as six spectral regions of the remnant. The spectrum of the whole SNR was fit by two thermal components: a high-temperature plasma ($\sim$1.4 keV) in NEI with an enhanced Si abundance and a low-temperature plasma ($\sim$0.8 keV) in ionization equilibrium. Unlike the whole SNR spectrum, the spectra from individual SNR regions were well fit with a single-temperature thermal model. The {\it Chandra} observations also revealed a compact central object inside the SNR shell, called CXOU J171801.0$-$372617. The X-ray morphology was found to be very similar to the radio morphology, thereby giving a hint of an expansion into a region with large density gradients as it is shown by the $^{12}$CO observations \citep{reynosomagnum2001, dubner2004}.

Five OH maser spots were found in the velocity range of 14.3 and 16.9 km s$^{-1}$ inside G349.7$+$0.2 showing clear evidence of interaction between the SNR and the molecular clouds (MCs) \citep{frail1996}. The magnetic field at the brightest maser is measured to be 350 $\pm$ 50 $\mu$G \citep{brogan2000}. The velocities of these OH maser spots are in agreement with the velocity of the shocked MC \citep{reynosomagnum2001, dubner2004} and the velocity of the far 3 kpc arm \citep{damethaddeus2008}.

Molecular line transitions found at similar velocities of those of the OH masers and the shock-excited near-infrared H$_2$ emission hints that the SNR-MC interaction is happening toward the center of the SNR \citep{lazendic2010,tianleahy2014}. By looking at the CO data and the results of the H\,{\sc i} absorption analysis, \citet{tianleahy2014} concluded that the clouds with a velocity of 16.5 km s$^{-1}$ are behind G349.7$+$0.2 and that they are shocked by the SNR, where the interaction gives rise to near-infrared H$_2$ emission and the OH maser emission.

It has been suggested that all maser-emitting (ME) SNRs could be of the mixed-morphology (MM) class \citep{rhopetre1998, yusefzadeh2003}.  It was shown by \citet{lazendic2005} that the X-ray morphology of G349.7$+$0.2 is not centrally peaked. It was suggested that this morphology can be explained by the projection effect model by \citet{petruk2001}, which assumes a shell-like evolution of an SNR at the edge of an MC, where the ambient density gradient does not lie in the projection plane. So it is highly likely that the G349.7$+$0.2 is expanding into a density gradient with an angle of $\sim$45$^{\circ}$, rather than expanding along the direction of the density gradient.

In gamma rays, MM SNRs interacting with MCs were among the first SNRs detected by the Large Area Telescope (LAT) on board {\it Fermi Gamma Ray Space Telescope} ({\it Fermi}-LAT). The gamma-ray luminosities of MM SNRs were found to be much higher than that of other detected SNRs, i.e. $\sim$10$^{35}$$-$10$^{36}$ erg s$^{-1}$ for IC443, W28, W51C, W44, W49B, and 3C391 \citep{abdoIC4432010,abdoW282010,abdoW51C2009,abdoW442010,abdoW49B2010,castroslane2010,ergin2014}. Interactions of these MM SNRs with MCs were clearly shown by the detection of 1720 MHz OH masers \citep{yusefzadeh1995, frail1996, green1997, claussen1997, hewitt2009} and near-infrared H$_2$ and [Fe\,{\sc ii}] lines \citep{keohane2007,reach2005}.

For G349.7$+$0.2, \citet{castroslane2010} reported a $\sim$10$\sigma$ detection of GeV gamma rays toward the position of G349.7$+$0.2 from the data collected by {\it Fermi}-LAT. No evidence of spatial extension was found and a power-law fit to the gamma-ray data gave a spectral index of $\Gamma$ = $-$2.10 $\pm$ 0.11. The shape of the gamma-ray spectrum showed a steepening of the spectral index after a few GeV. So, a fit to an exponential power-law function resulted in a slightly better fit  with a cut-off energy of $\sim$16.5 GeV and spectral index of $\Gamma$ = $-$1.74 $\pm$ 0.37. In TeV energies, \citet{abramowski2014} reported the detection of G349.7$+$0.2 at 6.6$\sigma$ and found the TeV flux at roughly 0.7\% of the Crab Nebula's gamma-ray flux with a spectral index of 2.8 $\pm$ 0.27$_{stat}$ $\pm$ 0.20$_{syst}$.

MM SNRs interacting with MCs are primary targets for the detection of gamma rays of hadronic origin \citep{slane2014}, where two gamma rays are produced from the decay of a neutral pion ($\pi^{\circ}$) created in a proton-proton interaction while the SNR shock passes through a dense molecular material. The gamma-ray spectra of these SNRs fall steeply below 250 MeV  and at energies greater than 1 GeV they trace the parent proton energy distribution \citep{ackermann2013}.

In addition to the fact that G349.7$+$0.2 is an ME SNR and shows a high gamma-ray luminosity ($\sim$10$^{35}$ erg s$^{-1}$) in the GeV energy range \citep{castroslane2010}, there are other clues which may support the MM-nature of G349.7$+$0.2, namely, the gamma-ray spectrum being dominated by the hadronic emission component and the X-ray spectrum presenting overionized (recombining) plasma features.

In recombining SNRs, the ionization temperature is higher than the electron temperature. This overionized plasma is a signature of rapid electron cooling. The existence of recombining plasma (RP) was discovered in IC443 and W49B during the X-ray studies of {\it ASCA} on six MM SNRs \citep{kawasaki2002, kawasaki2005}. Additionally, the X-ray Imaging Spectrometer, XIS, \citep{koyama2007} onboard {\it Suzaku} \citep{mitsuda2007} has discovered strong radiative recombination continuum (RRC) features from the following MM SNRs: IC443, W49B, G359.1-0.5, W28, W44, G346.6-0.2, 3C 391, and G290.1-0.8 \citep{yamaguchi2009, ozawa2009, ohnishi2011, sawada2012, uchida2012, yamauchi2013, ergin2014, kamitsukasa2014}.

Recently, \citet{yasumi2014} studied G349.7$+$0.2 using {\it Suzaku} data. They found that the spectrum of the whole SNR was well described by two optically thin thermal plasmas: a low-temperature ($\sim$0.6 keV) plasma in collisional ionization equilibrium (CIE) and a high-temperature ($\sim$1.24 keV) plasma in NEI. The low temperature plasma has solar metal abundances suggesting that the emission is dominated by the interstellar material. The high temperature plasma has super-solar abundances implying that the emission is influenced by the shocked ejecta. By using the abundance pattern of the ejecta component \citet{yasumi2014} estimated that the plasma originates from the core-collapse supernova (SN) explosion with the progenitor mass of $\sim$35$-$40 $M_{\sun}$ for this remnant. 

In this paper, we investigate the characteristics of the continuum radiation; thermal bremsstrahlung continuum and RRC in two regions, the brighter eastern (E) region and the western (W) region, by utilizing the superior spectral capabilities for diffuse sources of XIS onboard {\it Suzaku}. We report on our analysis of the E and W regions of G349.7$+$0.2. We analyze the GeV gamma-ray data, and probe the source morphology and variability of G349.7$+$0.2. In addition, we perform a detailed modeling of the GeV gamma-ray spectrum to understand if the gamma-ray emission from G349.7$+$0.2 might be the result of hadronic interactions between the SNR shock and the associated MC. 

\section{Data Analysis and Results}
\subsection{X-Rays} \label{x-rays}
\subsubsection{Observation and Data Reduction}
G349.7$+$0.2 was observed with {\it Suzaku} on 2011 September 19 with XIS for a total exposure time of 160 ks (observation ID: 506064010). The XIS consists of three active X-ray CCD detectors (XIS 0, 1, and 3). XIS0 and XIS3 have front-illuminated (FI) CCDs, while XIS1 contains a back-illuminated (BI) CCD.

For data reduction, we used the High Energy Astrophysics Software ({\sc HEASoft}) package version 6.16 and the calibration database (CALDB) version 20140520. We used {\sc xspec} version 12.8.2 \citep {Arnaud1996} for spectral analysis. The redistribution matrix and the auxiliary response functions were generated by \texttt{xisrmfgen} and \texttt{xissimarfgen} \citep {Ishisaki2007}, respectively.
\begin{figure}
\centering
  \vspace*{17pt}
  \includegraphics[width=10cm]{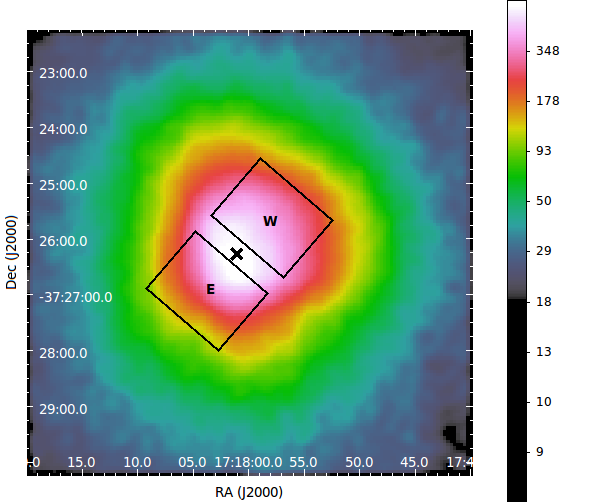}
  \caption{{\small {\it Suzaku} combined images of XIS0, XIS1, and XIS3 in the 0.3$-$10.0 keV energy band. The boxes are the spectral extraction regions corresponding to the E and W of the SNR. CXOUJ171801.0$-$372617 \citep {lazendic2005} is indicated as a black cross.}\vspace{0.3cm}}
 \label{figure_1}
\end{figure}
\subsubsection{Estimation of background} \label{Estimation of background}
Since G349.7$+$0.2 is located on the Galactic plane, the background of this SNR consists of the instrumental non-X-ray background (NXB), the Cosmic X-ray background (CXB), and the Galactic ridge X-ray emission (GRXE). Considering the position dependence of the GRXE, we constructed the background spectrum from the nearby blank sky (observation ID: 100028030) as the background is dominated by the GRXE \citep{uchiyama2013}.
The NXB spectra for the source and background were estimated from the {\it Suzaku} database of dark-Earth observations using the procedure of \cite{tawa2008}. The source and background spectra were made by subtracting the NXB spectra using \texttt{ mathpha}. NXB-subtracted background spectrum was subsequently subtracted from the source spectrum using {\sc xspec}.
\subsubsection{Spectral analysis} \label{Spectral analysis}
Figure \ref{figure_1} shows {\it Suzaku} combined XIS0, XIS1, and XIS3 image of G349.7$+$0.2 in the 0.3$-$10.0 keV energy band. In order to study the spectral properties, we extracted the data from a $1\farcm7\times1\farcm3$ size box centered at $\rm{R.A.}(2000)=17^{\rm{h}} 18^{\rm{m}} 03\fs7$ and $\rm{decl.}(2000)=-37\degr 26\arcmin 56\farcs2$ for the E region and from a same-size box centered at $\rm{R.A.}(2000)=17^{\rm{h}} 17^{\rm{m}} 57\farcs8$ and $\rm{decl.}(2000)=-37\degr 25\arcmin 37\farcs$9 for the W region. These regions are shown with black rectangles in Figure \ref{figure_1}. The E region corresponds to regions 1, 2, and 3 in Figure \ref{figure_1} of \citet{lazendic2005} and the W region includes the regions 4, 5, and 6 in the same figure.
\begin{figure}
  \vspace*{17pt}
  \includegraphics[width=0.50\textwidth]{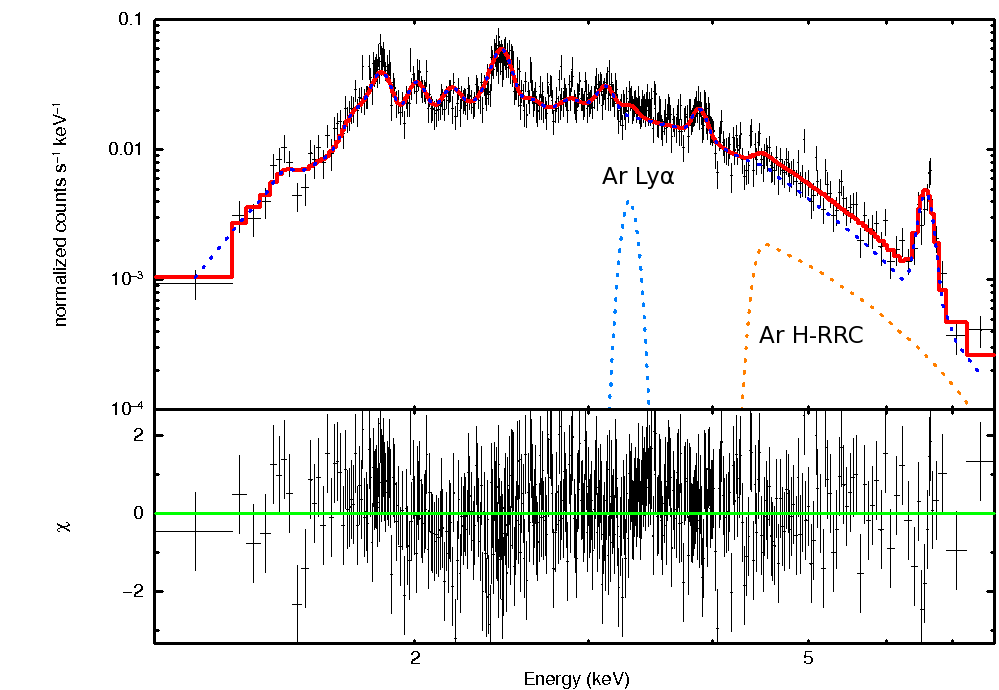}
  \includegraphics[width=0.5\textwidth]{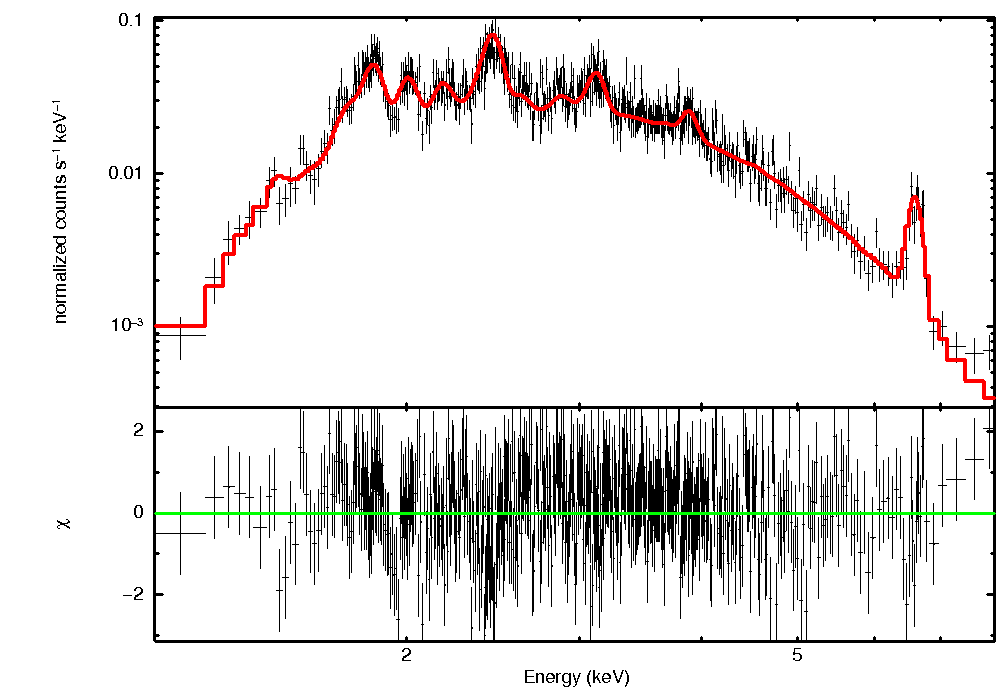}
  \caption{{\small {\it Suzaku} FI spectra shown in the 1.0$-$8.0 keV energy range with the absorbed VVPSHOCK plus RRC model and the gaussian line for the E region (upper panel) and the VVPSHOCK model for the W region (bottom panel).}\vspace{0.3cm}}
\label{figure_2}
\end{figure}

\begin{table*}
\caption{\small{Best-fit parameters with corresponding errors at 90\% confidence level for the E and W spectra of G349.7$+$0.2. The spectral fits were carried out in the 1.0$-$8.0 keV energy band.}}
 \begin{minipage}{170mm}
 \begin{center}
   \begin{tabular}{cccc}
  \hline\hline
Component                   &Parameters                                           & E                 & W \\
\hline
TBABS                        &N$_{\rm{H}}$ (10$^{22}$ cm$^{-2}$)     & 8.7$_{-0.5}^{+0.3}$       &9.5$\pm0.3$\\
VVPSHOCK              &kT$_{e}$ (keV)                         &1.2$\pm0.1$     & 1.3$\pm0.1$\\
                                    &Si (solar)                             &1.4$_{-0.2}^{+0.3}$   &1.2$\pm0.1$\\
                                    &Ar (solar)                              &1 (fixed)  &1.2$\pm0.2$   \\
                                    &Ca (solar)                         &2.0$_{-0.5}^{+0.7}$&1.3$_{-0.3}^{+0.4}$\\
                                    &Fe (solar)                             &3.3$_{-1.5}^{+2.1}$       & 1.5$\pm0.4$\\
                                    &$\tau_{\rm u}$ (10$^{11}$ cm$^{-3}$ s)                 &13.9$_{-6.3}^{+1.7}$     &4.8$_{-0.9}^{+1.2}$\\
                                        &Norm (10$^{-2}$photon cm$^{-2}$ s$^{-1}$)           &5.2$_{-0.7}^{+0.4}$         &5.8$\pm0.3$\\
\hline
Ar Ly$\alpha$          &E (keV)                                &3.3 (fixed)        & \\
                                   &$\sigma$ (keV)                     &0.0 (fixed)    & \\
                                   &Norm (10$^{-5}$photon cm$^{-2}$ s$^{-1}$)  &1.4$_{-0.8}^{+0.7}$     & \\
\hline
RRC H-like Ar                  & E (keV)                               & 4.4 (fixed)       &\\
                            &Norm (10$^{-5}$photon cm$^{-2}$ s$^{-1}$)  &3.0$_{-1.5}^{+0.9}$&\\
\hline
                            &$\chi^2$/dof                           &432/427            &539.7/529\\
                            &reduced $\chi^2$                       &1.01                &1.02\\
\hline
\end{tabular}
  \label{table_1}
  \end{center}
\end{minipage}
\end{table*}

For the E region, we first tried to fit the spectrum to a single-temperature plane-parallel shock model with variable abundances for all elements ($Z\le30$) (VVPSHOCK in {\sc xspec}) modified by {\sc TBABS} absorption model \citep{wilms2000} in {\sc xspec}. The abundances are set to \texttt{wilm} \citep{wilms2000}. In this fitting, the absorbing column density ($N_{\rm H}$), upper limit of the ionization timescale ($\tau_{\rm u}$), electron temperature ($kT_{\rm e}$), normalization, and abundances of Si, Ca, and Fe were left as free parameters. The other elements were fixed at their solar values \citep{wilms2000}. Although this fit gave an acceptable $\chi^{2}$/degrees of freedom (dof) value of 448/429, residuals at $\sim$3.3 keV and $\sim$4.1$-$4.5 keV remained. The first residual around 3.3 keV corresponds to Ar Ly$\alpha$ line. The other residual could be caused by the RRC corresponding to the H-like Ar.

Thus we added a Gaussian and RRC (REDGE in {\sc xspec}) components to the model. The line energy of the Ar Ly$\alpha$ was fixed at 3.3 keV, while the normalization was a free parameter. We note that the Gaussian line width parameter was fixed to zero eV. For the RRC component, the energy was fixed at 4.4 keV. After adding the Gaussian and RRC components to the model, the reduced $\chi^{2}$ value improved to 1.01 for 427 dof, which yields an {\it F}-test probability of 4.2 $\times$ 10$^{-4}$.  

We also applied RP model VVRNEI in {\sc xspec} for E region. VVRNEI is a non-equilibrium recombining collisional plasma model with variable abundances for all elements ($Z\le30$). It is characterized by a constant electron temperature ($kT_{\rm e}$), initial temperature ($kT_{\rm init}$), and a single ionization parameter. N$_{\rm{H}}$, $kT_{\rm e}$, the abundances of Si, Ca, Fe, $\tau$, and the normalization were free parameters during the fitting. We tried fitting by fixing $kT_{\rm init}$ to a range of values (2.0, 5.0, 10.0, and 50.0 keV) and for each fit we obtained similar $\chi^2$/dof and $\tau$ values for each of the $kT_{\rm init}$ values. This model gave a $\chi^2$/dof of 520.6/431, which required a large ionization timescale ($\tau$ $\sim$ 2.8 $\times$ 10$^{13}$ cm$^{-3}$ s) showing that G349.7$+$0.2 is in ionization equilibrium and not in RP phase.

As a next step, we applied an absorbed VVPSHOCK model for the W region. During the fitting, the absorption, electron temperature, ionization timescale, abundances of Si, Ar, Ca, and Fe, and normalization parameters were allowed to vary freely, while the other elements were fixed at their solar values \citep{wilms2000}. We obtained an acceptable fit with a $\chi^{2}$/dof value of 539.7/529 and found no significant residuals. So, the spectral analysis of the E and W regions showed that there is no overionized plasma in G349.7$+$0.2. The spectral fits for the E and W regions are given in Figure \ref{figure_2} and the best-fit parameters are summarized in Table \ref{table_1}, where all the errors are given at 90\% confidence level.

\subsection{ Gamma Rays} \label{gamma}
In this work, {\it Fermi}-LAT data from 2008 August 4 to 2014 March 27 were analyzed. The events-data were selected from a circular region of interest (ROI) with a radius of 18$^{\circ}$ centered at the SNR position of $\rm{R.A.~}(J2000) = 17^{\rm{h}} 17^{\rm{m}} 59\fs4$, $\rm{decl.~}(J2000)=-37\degr 25\arcmin 59\farcs88~(\rm{R.A.} = 259^{\circ}\!.496~\rm{and}~\rm{decl.} = -37^{\circ}\!.433)$. Using \emph{gtselect} of Fermi Science Tools (FST), we chose the {\it Fermi}-LAT Pass 7 events suggested for the galactic point source analysis. We only selected the events coming from zenith angles smaller than 105$^{\circ}$, to reduce the contributions from the albedo gamma rays from the Earth's limb.

\begin{figure}
\centering
\includegraphics[width=0.5\textwidth]{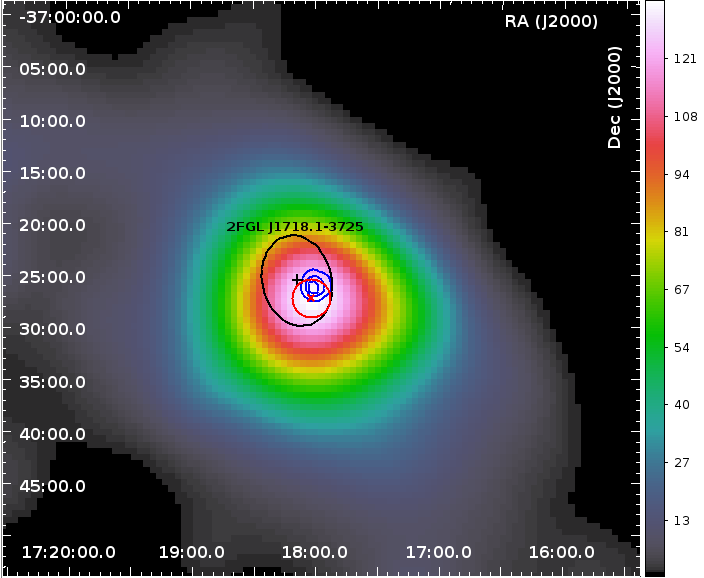}
\caption{ {\small Gamma-ray TS map of G349.7$+$0.2 and its neighborhood with a bin size of 0$^{\circ}\!$.01 $\times$ 0$^{\circ}\!$.01. The black cross and circle are representing the GeV source location and its positional statistical error from the 2nd {\it Fermi}-LAT catalog corresponding to G349.7$+$0.2, respectively. The red cross and circle is the best-fit location and its statistical error circle  (95\% confidence level) found in this analysis, respectively. The blue contours show the {\it Suzaku} data from Figure \ref{figure_1}, where the 3 contours represent X-ray counts higher than 190, 365, and 540 counts (cts).} \vspace{0.2cm} }
\label{figure_3}
\end{figure}
To study G349.7$+$0.2's spatial and spectral characteristics, the maximum likelihood fitting method \citep{mattox1996} was employed on the spatially and spectrally binned data within a square region of $\sim$20$^{\circ}\!$ $\times$ 20$^{\circ}\!$ using the P7SOURCE$_{-}\!$V6 version of the instrument response function. The analysis was performed using \emph{pointlike} \citep{kerr2011, lande2012} (FST-v9r32p0) and the standard binned likelihood analysis (FST-v9r27p1) packages, both based on \emph{gtlike} \citep{abdo2009}, for cross-checking the validity of the results.

The model of the analysis region contains the diffuse background sources and all the point-like sources from the 2nd {\it Fermi}-LAT source catalog located within a distance of 18$^{\circ}$ from the ROI center. We fixed all parameters of the point-like sources in the model, except 2 sources within the distance of 2$^{\circ}$ from the ROI center, where we set the normalization and spectral parameters free. The standard diffuse background model has two components: the diffuse Galactic emission (\emph{gal$_{-}$2yearp7v6$_{-}$v0.fits}) and the isotropic component (\emph{iso$_{-}$p7v6source.txt}), which is a sum of the extragalactic background, unresolved sources, and instrumental background. The normalization of the isotropic component was kept fixed, while the normalization of the galactic diffuse background was set free during the analysis.

The test statistics (TS) parameter having larger values indicates that the null hypothesis (maximum likelihood value for a model without an additional source) is incorrect, where the detection significance of a source can be approximated to the square root of the TS value.

\subsubsection{Location and Spectrum}
We searched for the best-fit location of G349.7$+$0.2 within the ROI, which was found as $\rm{R.A.~(J2000)} = 259^{\circ}\!.509~\pm~0^{\circ}\!.013~\rm{and}~\rm{decl.~(J2000)} = -37^{\circ}\!.455~\pm~0^{\circ}\!.013$. The TS value of the best-fit position\footnote{
$\rm{R.A.~}(J2000) = 17^{\rm{h}} 18^{\rm{m}} 02\fs16$, $\rm{decl.~}(J2000)=-37\degr 27\arcmin 18\farcs00$} was enhanced by 2.7$\sigma$ over the position of 2FGL J1718.1$-$3725 in the 2nd {\it Fermi}-LAT catalog. Then the model was refitted using the best-fit position to compute the TS map (Section \ref{exte}) and the spectrum.

To check the functional form of the spectral energy distribution (SED), we considered G349.7$+$0.2 as a point-like source. First, the power-law (PL) function was fit to the data between 200 MeV and 300 GeV. Because the spectrum deviates from a PL function, we tested if the gamma-ray emission is better described by a log-parabola (LP) or a broken power-law (BPL) function, where their functional forms are as follows:
\begin{itemize}
\item Log-parabola: \\
F(E)$^{LP}$ =  N$_{\circ}$ (E/E$_b$)$^{(\Gamma_1 + \Gamma_2 \mbox{{\small ln(E/E$_b$)}})}$
\item Broken Power-law:\\
F(E)$^{BPL}$ =  N$_{\circ}$ (E/E$_b$)$^{-\Gamma_1}$ for  E $<$ E$_b$ \\
$~~~~~~~~~~~~~$ =  N$_{\circ}$ (E/E$_b$)$^{-\Gamma_2}$ for  E $>$ E$_b$
\end{itemize}

Here, F(E)$^{LP}$ and  F(E)$^{BPL}$ are representing the flux for the LP and BPL type spectrum, respectively. The spectrum breaks or turns at a certain energy of E$_b$, where the spectral index changes from $\Gamma_1$ to $\Gamma_2$. N$_{\circ}$ is the normalization parameter.

Using different functions in fitting the spectrum of G349.7$+$0.2, the likelihood ratio, TS, was used as a measure of the improvement of the likelihood fit with respect to the simple PL. The results to the spectral fits are summarized in Table \ref{table_3}, where the highest TS value (168) was found for the BPL fit.

The PL resulted in spectral index of $\Gamma_1$ = 2.00 $\pm$ 0.03, which is in agreement with the best-fit power-law index value given for 2FGL J1718.1$-$3725 in the 2nd {\it Fermi}-LAT catalog ($\sim$1.98), \citet{nolan2012}. This result also matches to the results obtained by \citet{castroslane2010}. The best-fit parameters for the BPL fit are N$_{\circ}$ = (2.30 $\pm$ 0.21) $\times$ 10$^{-11}$  MeV$^{-1}$ cm$^{-2}$ s$^{-1}$, $\Gamma_1$ $\sim$ 4.98, and $\Gamma_2$ =  2.28 $\pm$ 0.04, where the given uncertainties are statistical. The total energy flux was found as (3.70 $\pm$ 0.10) $\times$ 10$^{-11}$ erg  cm$^{-2}$ s$^{-1}$.

\begin{table}
\caption{{\small Spectral fit parameters for PL, LP, and BPL between 200 MeV and 300 GeV assuming G349.7$+$0.2 as a point-like source.}}
\scalebox{0.87}{
\begin{tabular}{|l|c|c|c|c|c|}
\hline Spectral  & Photon Flux               & $\Gamma_1$            & $\Gamma_2$     & E$_b$              & TS       \\
           Model      & [ 10$^{-8}$ ph cm$^{-2}$ s$^{-1}$]  & $~$  & $~$                       & [MeV]                & $~$     \\\hline
           PL            & 3.59 $\pm$ 0.21        & 2.00 $\pm$ 0.03      & $-$                         & $-$                    & 137   \\
           LP            & 1.66 $\pm$ 0.13        & 1.67                            & 0.19 $\pm$ 0.31  & 1310                & 151        \\
          BPL          & 1.24 $\pm$ 0.14        & 4.98                            & 2.28 $\pm$ 0.04  & 568                  & 168      \\\hline
\end{tabular}
}
\label{table_3}
\end{table}

\begin{figure}[t]
\centering
\includegraphics[width=0.5\textwidth]{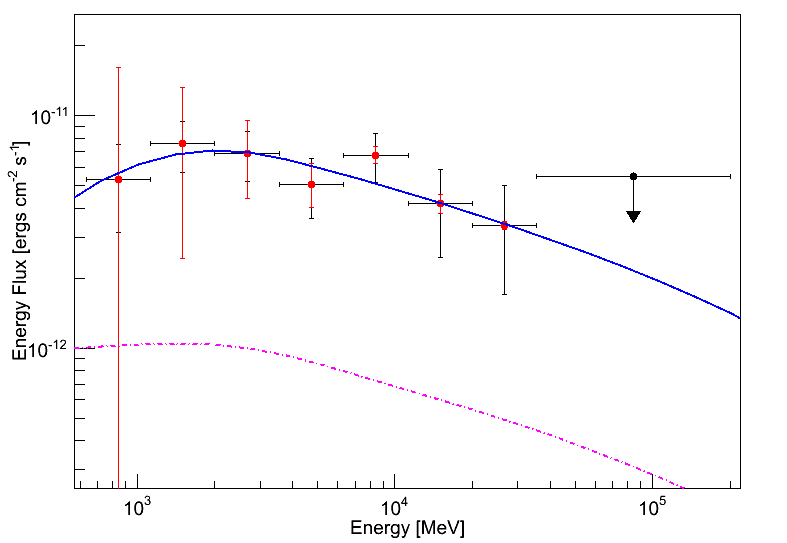}
\caption{\small{ The gamma-ray spectrum of G349.7$+$0.2. The {\it Fermi}-LAT spectral data points are represented by the red filled circles with their corresponding statistical and systematic errors shown in black and red color, respectively. The models of emission are shown by the thick blue ($\pi^{\circ}$-decay component) and the dashed-dotted magenta (the bremsstrahlung emission component) lines. The parameters used to estimate the  emission spectra are explained in Section \ref{modeling}.} \vspace{0.3cm} }
\label{figure_4}
\end{figure}
We also calculated the systematic errors that stem from the uncertainties in the Galactic diffuse background intensity, where we followed the methods described in \cite{abdoW51C2009} and \cite{castro2013}. We varied the normalization value of the Galactic background by $\pm$6\% from the best-fit value and used these new frozen values of the normalization parameter to recalculate the SED of G349.7$+$0.2. The systematic errors on the SED are shown in Figure \ref{figure_4} with red solid lines on top of the statistical errors given in black lines.
\subsubsection{Extension \label{exte}}
To investigate the morphology of G349.7$+$0.2, we created a 2$^{\circ}$ $\times$ 2$^{\circ}$ TS map of G349.7$+$0.2 and its neighborhood with a bin size of 0$^{\circ}\!$.01 $\times$ 0$^{\circ}\!$.01. The TS map shown in Figure \ref{figure_3} was produced with {\it pointlike} using a background model file, which contained all the point-like sources and diffuse sources, but excluded G349.7$+$0.2 from the model. So, it shows the TS distribution of gamma rays originating dominantly from G349.7$+$0.2. In Figure \ref{figure_3} the blue contours represent the {\it Suzaku} combined XIS image in the 0.3$-$10.0 keV energy band (from Figure \ref{figure_1}) and the black cross and circle represent the location and its statistical error of the GeV source from the 2nd {\it Fermi}-LAT catalog corresponding to G349.7$+$0.2, respectively. The best-fit location of the gamma-ray emission is slightly offset from the center of the X-ray remnant, as shown in Figure \ref{figure_3} by the red cross and circle, and it is closer to the south-east region of the SNR.

To search for the energy dependent morphology, we split the data set into two energy ranges (200 MeV$-$1 GeV and 1$-$300 GeV) and analyzed the data in each range. We found no significant gamma-ray excess at the location of G349.7$+$0.2  for the energy range between 200 MeV and 1 GeV, but G349.7$+$0.2 was detected in the higher-energy range of 1$-$300 GeV with a significance of $\sim$11$\sigma$ using the BPL spectral model.

Assuming that G349.7$+$0.2 has a PL/BPL type spectrum, we checked the extension of G349.7$+$0.2 using the TS of the extension (TS$_{ext}$) parameter.  TS$_{ext}$ is the the likelihood ratio comparing the likelihood for being a point-like source to a likelihood for an existing extension. The {\it pointlike} code calculates the TS$_{ext}$ by fitting a source first with a disk template and then as a point-like source. In {\it pointlike}, the extended source detection threshold is defined as the source flux, where the TS$_{ext}$ value averaged over many statistical realizations is 16 \citep{lande2012}. Simulation studies showed that to resolve a disk-like extension with a radius (r) of 0$^{\circ}\!$.1 at the detection threshold, the source with a spectral index value of 2.0 and 2.5 must have a minimum flux of 3 $\times$ 10$^{-7}$ ph cm$^{-2}$ s$^{-1}$ and 2 $\times$ 10$^{-6}$ ph cm$^{-2}$ s$^{-1}$, respectively \citep{lande2012}.

\begin{table}[h]
\caption{{\small $\!\!\!$Fit results of disk-like extension model applied to G349.7$+$0.2 gamma-ray data.} }
\centering
\begin{tabular}{|l|c|c|c|}
\hline Spectral                & Longitude                    & Latitude                       & Sigma               \\
           Model                    & [$^{\circ}$]                   & [$^{\circ}$]                   & [$^{\circ}$]        \\ \hline
           PL                          & 349.70 $\pm$ 0.01    & 0.158 $\pm$ 0.013    & 0.003                \\
           BPL                       & 349.70 $\pm$ 0.01     & 0.154 $\pm$ 0.013   & 0.005                 \\ \hline
\end{tabular}
\label{table_4}
\end{table}

For G349.7$+$0.2 having either a PL or a BPL type spectrum, the TS$_{ext}$ value was found to be about zero, after a disk template fitting, where the fit parameters are shown in Table \ref{table_4}. This indicates that a disk-like extension with r $\sim$ 0$^{\circ}\!$.1 could not be resolved at the integrated flux level of 3.59 $\times$ 10$^{-8}$  ph cm$^{-2}$ s$^{-1}$ for a  PL type spectrum with a spectral index of 2.0 and of 1.24 $\times$ 10$^{-8}$  ph cm$^{-2}$ s$^{-1}$ for the BPL type spectrum with a spectral index of 2.3 ($\Gamma_{2}$), respectively.

\subsubsection{Variability and Pulsation}
\begin{figure}[h]
\begin{center}
\includegraphics[width=0.5\textwidth]{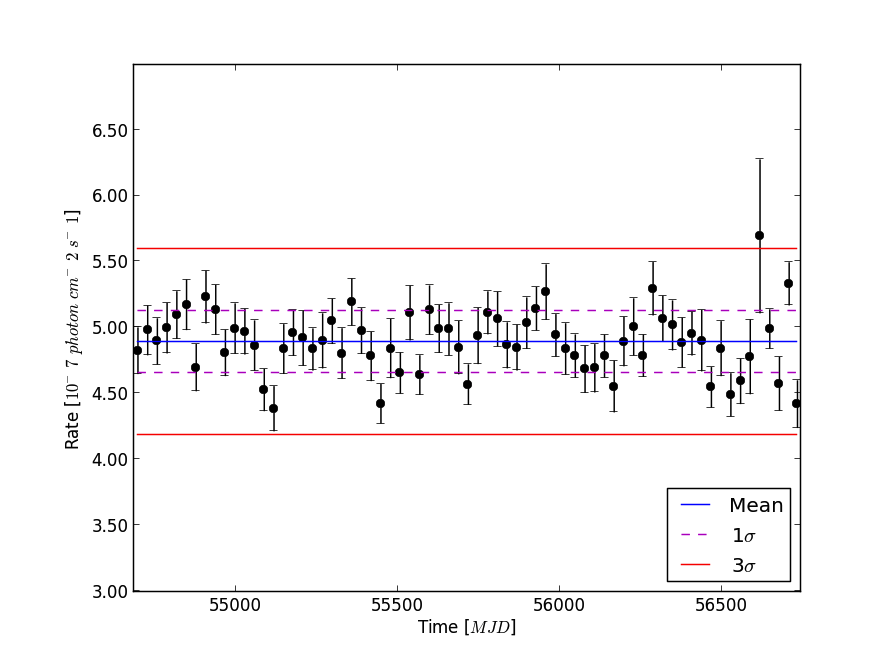}
\caption{\small{The 1-month binned light curve showing no significant variability in G349.7$+$0.2. }}
\label{figure_5}
\end{center}
\end{figure}

We first looked for long term variability in the light curve of G349.7$+$0.2 by taking data from the circular region of 1$^{\circ}$ around the best-fit position.
Figure \ref{figure_5} shows the 1-month binned light curve obtained after applying {\it Fermi}-LAT {\it aperture photometry}, where we checked for possible variations in the flux levels. 
If any or some of the flux data points are above 3$\sigma$, this would be an indication of a significant variability (e.g. flare) in the circular region of interest. 
In Figure \ref{figure_5}, most of the flux data points remain within the 1$\sigma$ and 3$\sigma$ bands. Although one flux data point is above 3$\sigma$, we note that it's error bars are large, showing a large statistical uncertainty in this time interval. Thus, we conclude that there is no long term variability observed in the close neighborhood of G349.7$+$0.2.

We have also checked if the spectral shape of G349.7$+$0.2 fits to the standard spectrum of a pulsar, Power Law with Exponential Cutoff (PLEC). The best-fit cutoff energy was found as 24.60 $\pm$ 4.95 GeV, which is an order of magnitude away from the range of typical pulsar cutoff energies \citep{abdopulsar2010}. The PLEC fit did not show a significant improvement over the PL, BPL, and LP spectral fits.

\subsubsection{Molecular Environment  \label{molen}}
We used the CO data taken by the 12 m telescope of the NRAO \citep{reynosomagnum2001} with a good angular resolution (54$''$) to understand the molecular environment around G349.7$+$0.2. We integrated the CO spectrum over the whole range of velocities from $-$137 to $+$57 km s$^{-1}$ to obtain the velocity integrated CO intensity (W$_{CO}$). The W$_{CO}$ averaged over the region with a radius of 0$^{\circ}\!$.025 covering the X-ray remnant was found to be $\sim$150 K km s$^{-1}$. Using the CO-to-H$_2$ conversion factor of X = 1.8 $\times$ 10$^{20}$ cm$^{-2}$ K$^{-1}$ km s$^{-1}$ \citep{dame2001}, we found N(H$_2$) = 2.7 $\times$ 10$^{22}$ cm$^{-2}$. Using the N(H\,{\sc i}) value found by \citet{tianleahy2014}, we calculated the total column density as N(total) = N(H\,{\sc i}) $+$ 2N(H$_2$) = 7.12 $\times$ 10$^{22}$ cm$^{-2}$. \citet{lazendic2005} calculated the total column density as 7.3 $\times$ 10$^{22}$ cm$^{-2}$ adding the contributions from H\,{\sc i}, 2.8 $\times$ 10$^{22}$ cm$^{-2}$ assuming T$_s$ = 140 K, and H$_2$,  4.6 $\times$ 10$^{22}$ cm$^{-2}$.

\begin{figure}
\centering
\includegraphics[width=0.50\textwidth]{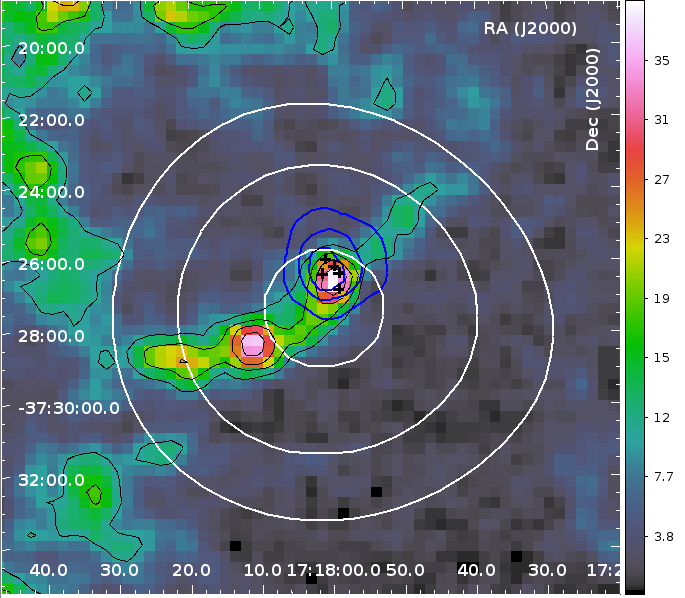}
\caption{\small{The CO data \citep{reynosomagnum2001} integrated in the +14 to +20 km s$^{-1}$ velocity range around the G349.7$+$0.2 location, where the black contours are for W$_{CO}$ at 10, 18, 24, and 31 K km s$^{-1}$. The white contours represent the gamma-ray TS values of 90, 110, and 130. The blue contours show the X-ray counts from {\it Suzaku}, which are at 190, 365, and 540 cts. The black plus markers are the five masers found by \citet{frail1996}. } \vspace{0.3cm}}
\label{figure_6}
\end{figure}

Figure \ref{figure_6} shows the CO map produced in the velocity range of [$+$14, $+$20] km s$^{-1}$, where white contours representing the TS values shown in Figure \ref{figure_3}. Also shown with black plus markers are the five masers measured by \citet{frail1996}, which are located inside the blue {\it Suzaku} X-ray contours of 540 cts.

The progenitor of the G349.7$+$0.2 has exploded within a large ringlike CO shell (seen as the color-coded arc structure in Figure \ref{figure_6}), a remnant of an older SN explosion containing denser clumps of MCs. The explosion probably happened within the inter-clump gas \citep{reynosomagnum2001} of the CO shell, at a location between the edge of the CO shell and an MC clump inside this shell. In our interpretation we assume that while the SNR's shell was expanding, a small section of it broke out of the CO shell boundary and entered the rarefied interstellar medium (ISM) moving in a direction toward the observer. 

The X-ray morphology of G349.7$+$0.2 is similar to the radio image consisting of a two-ring structure, where one ring is smaller and brighter than the other. The smaller ring might be the part of the SNR shock front entering into a higher density region, an MC clump called as `cloud1' by \citet{reynosomagnum2001}, and interacting with it. The shock of G349.7$+$0.2 entering `cloud1' was confirmed to be of type C, which allows OH 1720 MHz masers to form \citep{lockett1999}. The larger ring structure represents the SNR shock expanding into the relatively less denser medium of the CO shell. 

Therefore, we calculated the proton density for two regions: one having a radius of 0$^{\circ}\!$.015 corresponding to the inner smaller ring, and another one with a radius of 0$^{\circ}\!$.022 representing the larger outer ring. The smaller ring radius was chosen such that it covers the highest number of X-ray counts (shown with the innermost blue contour for $>$540 cts in Figure \ref{figure_6}). This region coincides mostly with the E region of the SNR selected in the X-ray analysis (Section 2.1.3). The larger ring was chosen to include the X-ray contour representing the X-ray counts higher than 365, where the W region is also partially covered. Both rings are within the 10$\sigma$ contours of GeV gamma-ray TS distribution. To calculate the average proton density within the regions surrounded by these two rings, we first estimated the size of the emission regions to be $\sim$3 and $\sim$4.5 pc using the distance to the SNR to be 11.5 kpc. Then assuming a spherical geometry of the emission region, the proton density (n) corresponding to the inner and outer ring region was found to be 81 and 35 cm$^{-3}$, respectively.

\subsubsection{Modeling and Interpretation \label{modeling}}
To understand whether the leptonic or hadronic scenario is dominating the SED of G349.7$+$0.2, we started the multi-wavelength modeling assuming a BPL type of spectra for both the electron and proton distributions inside the source:

\begin{eqnarray}
{dN_{e,p} \over dE}~& = & ~N^1_{e,p} ~E_{e,p}^{-\alpha} ~\mbox{for} ~E_{e,p} <E_{br} \nonumber \\
 &=&  ~N^2_{e,p}~ E_{e,p}^{-\beta}~ exp{\left(-{E_{e,p} \over E_c} \right )} ~ \mbox{for}~ E_{br}\leq E_{e,p} \leq E_c ~~~~~~
\label{eqn:pi0}
\end{eqnarray}
where E$_{e,p}$ is the electron or proton energy and E$_{br}$ is the spectral break energy with the spectral index changing from $\alpha$ to $\beta$. E$_c$ is the maximum energy of electrons or protons. N$^1_{e,p}$ and N$^2_{e,p}$ are normalization constants.

\begin{figure}
\includegraphics[width=0.5\textwidth]{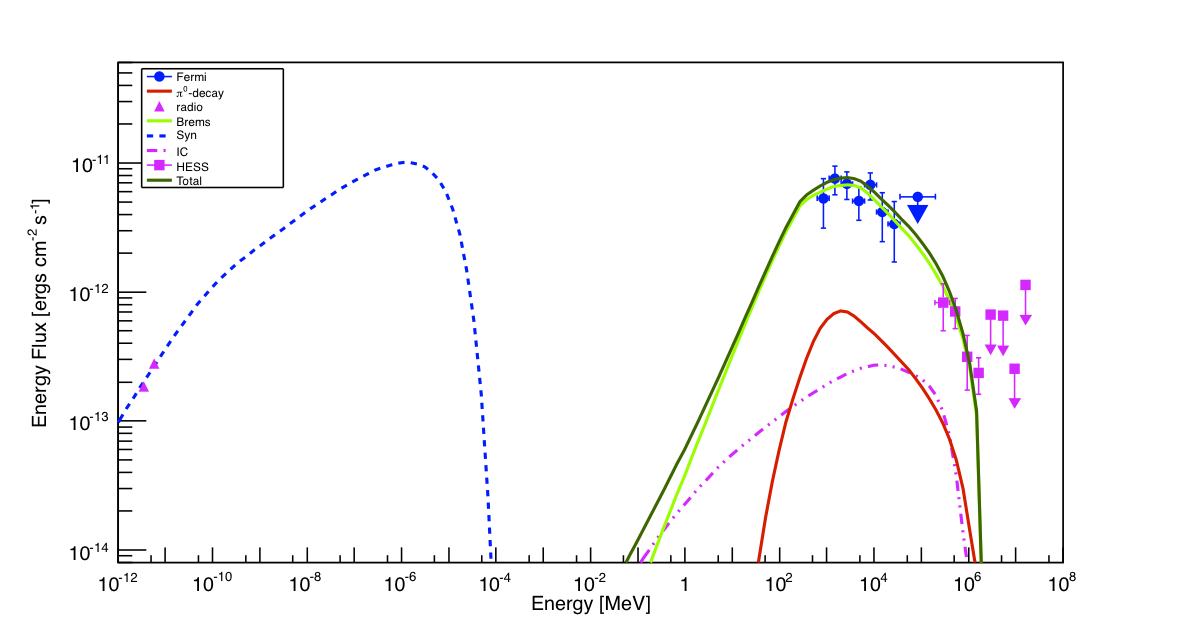}
\includegraphics[width=0.5\textwidth]{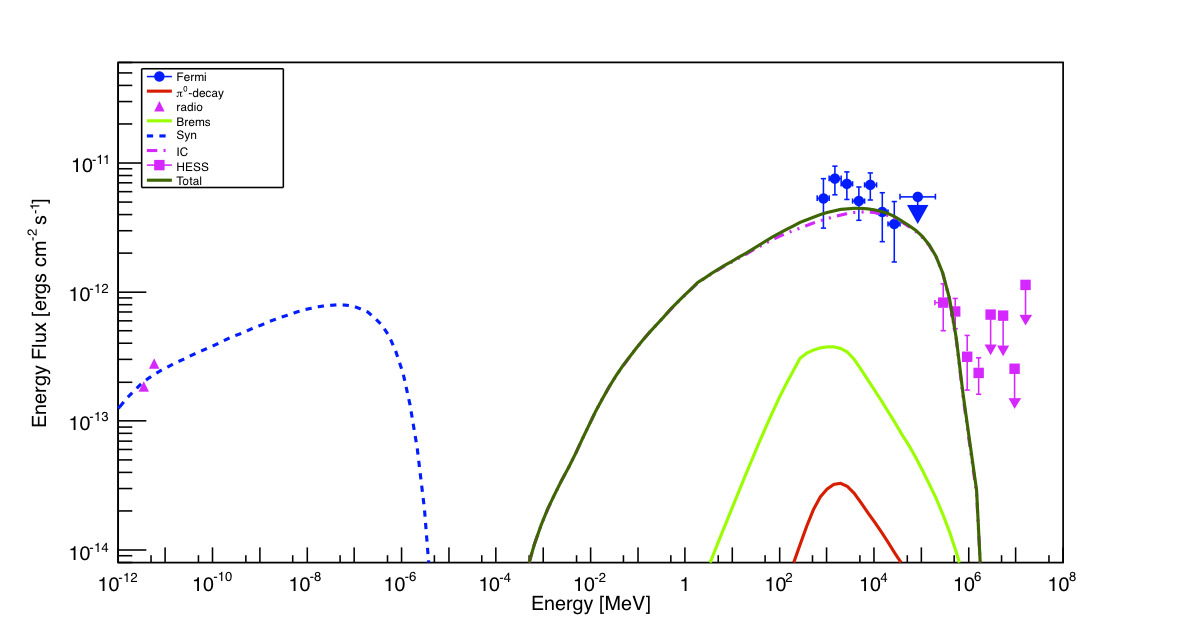}
\includegraphics[width=0.5\textwidth]{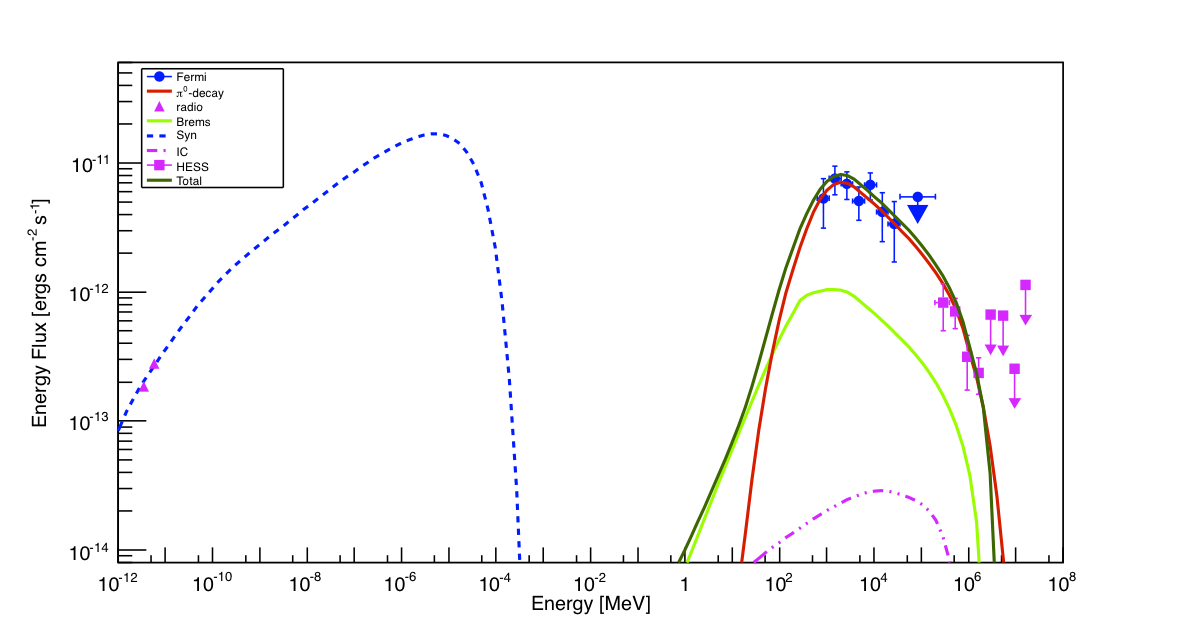}
\caption{\small{The different emission models fit on the multi-waveband data of G349.7$+$0.2 for the proton density value of 81 cm$^{-3}$. The radio data  \citep{whiteoakgreen1996, green2009} given in magenta filled triangles, are fit by the synchrotron emission model shown in dashed blue line. {\it Fermi}-LAT data points are represented by the blue filled circles and their corresponding statistical errors are also shown in blue. The H.E.S.S. results \citep{abramowski2014} are shown by  the magenta filled squares. The solid light green and the double-dot-dashed magenta lines are the spectra corresponding to the bremsstrahlung and the IC models, while the solid red line shows the $\pi^{\circ}\!$-decay contribution. The total emission is given in solid dark-green line. The top, middle, and bottom panel shows the bremsstrahlung, IC, and the $\pi^{\circ}\!$-decay dominated model, respectively.} \vspace{0.3cm}}
\label{figure_7}
\end{figure}

For the multi-wavelength modeling we used the radio data \citep{shaver1985, whiteoakgreen1996, green2009} and the gamma-ray data from {\it Fermi}-LAT and H.E.S.S. The H.E.S.S. flux data points \citep{abramowski2014} are shown in  magenta color in Figure \ref{figure_7}. During the fitting procedure we took E$_c$ as 10 TeV for protons and 2 TeV for electrons. The value of E$_c$ was chosen such that it would not cause the spectra of bremsstrahlung and inverse-Compton (IC) overestimate the TeV fluxes, when they are considered the dominant models for explaining the data. Similarly, E$_c$ for protons was chosen such that the hadronic model does not overestimate the TeV fluxes.

We first considered a purely leptonic model to explain the observed fluxes from radio wavelengths to gamma rays. In the case of leptonic model, bremsstrahlung and IC emission processes \citep{blumenthalgould1970} were used for explaining the GeV$-$TeV data. The total energetics of the electrons and protons need to be known to estimate the electron to proton ratio, defined as the ratio of dNe/dE and dNp/dE at E = 10 GeV, and the different model contributions. We first estimated the bremsstrahlung spectrum using a fixed value of ambient proton density. Parameters of the leptonic model then were adjusted in such a way that the  bremsstrahlung spectrum could explain the observed fluxes at GeV$-$TeV energies. These parameters then were fixed and they were used in computing the IC and synchrotron spectra of the leptonic model. To explain the observed radio fluxes, the magnetic field was adjusted to get radio synchrotron spectrum. Using the same ambient proton density we estimated the hadronic contribution to gamma-ray spectrum resulting from the decay of $\pi^{\circ}$ \citep{kelner2006}. Secondly, we repeated a similar procedure for the case of IC process as the dominant model explaining the observed fluxes at GeV$-$TeV energies. And at last, we considered the $\pi^{\circ}\!$-decay model dominating among all the processes, which is a purely hadronic model \citep{kelner2006}. The parameters used for all three dominating processes are shown in Table \ref{table_5} for the two ambient proton densities of 35 and 81 cm$^{-3}$ derived in Section \ref{molen}.

\begin{table}
\caption{Parameters of multi-wavelength model. For bremsstrahlung and $\pi^{\circ}\!$-decay dominated models ambient proton density is considered to be 35 cm$^{-3}$ and 81 cm$^{-3}$.}
\scalebox{0.80}{
\begin{tabular}{|c|c|c|c|c|c|c|c|c|c|}
\hline
\multicolumn{1}{|c|}{Model} &\multicolumn{5}{|c|}{Parameters}&\multicolumn{2}{|c|}{Energetics} \\
\hline
                                & $\alpha$ & $\beta$ & $k_e/k_p$ & n                      & B                & Leptonic                &Hadronic \\
                                &                  &                &                     &  ($cm^{-3}$)  & ($\mu$G)  & ($10^{49}$ ergs) & ($10^{49}$ ergs)\\
\hline
\multicolumn{8}{c}{For n = 35 cm$^{-3}$}\\
\hline
Bremsstrahlung     & 1.8       & 2.4       & 1.0       &  35       & 20       & 1.25                 & 0.74\\
IC                              & 1.8       & 2.6       & 1.0       &  0.1      & 3.0      & 27.5                & 13.1\\
$\pi^{\circ}$-decay & 1.9      & 2.4       & 0.01     &  35       & 100     & 0.13                & 7.44\\
\hline
\multicolumn{8}{c}{For n = 81 cm$^{-3}$}\\
\hline
Bremsstrahlung      & 1.8       & 2.4       & 1.0       &  81       & 40       & 0.55               & 0.31\\
IC                               & 1.8       & 2.6       & 1.0       &  0.1      & 3         & 27.5               & 13.1\\
$\pi^{\circ}$-decay  & 1.9       & 2.4      & 0.01     &  81       & 150     & 0.07               & 3.13\\
\hline
\end{tabular}
}
\label{table_5}
\end{table}

Since the ambient proton density of 81 cm$^{-3}$ represents the interaction region corresponding to the smaller and brighter SNR shell, including the five OH masers, the modeling  results for this density will be important in terms of giving us clues about the proton spectrum. The best-fit parameters for the proton spectrum were obtained by a $\chi^2$-fitting procedure to the flux points. The dominating models of bremsstrahlung, IC, and $\pi^{\circ}$-decay for the ambient proton density of 81 cm$^{-3}$ are shown in the top, middle, and bottom panels of Figure \ref{figure_7}, respectively.  The estimated parameters are $\alpha~ = ~1.89 ~\pm~ 0.14$, $\beta ~ = ~2.42 ~\pm~ 0.03$, and $E_{br}$ = 12 GeV. The $\chi^2$/dof is estimated to be $\simeq$ 0.53. The best-fit gamma-ray spectrum resulting from the decay of $\pi^{\circ}$ is shown in the bottom panel of Figure \ref{figure_7} with the red solid line. The estimated total energy can be written as W$_{p}$  $\simeq$ 3.13 $\times$ 10$^{49} \left({81 ~\mbox{cm}^{-3} / n}\right)$ erg, where $n$ is the effective gas number density for p-p collision.

\vspace{0.5cm}

\section{Discussion}
\label{section:discussion}
G349.7$+$0.2 is an example of an ME SNR having an irregular shell that lacks central cavity in radio wavelengths \citep{shaver1985}, where the radio morphology is very similar to the one in X-rays. In both wavelengths the E side of the SNR is brighter. The most likely explanation for G349.7+0.2 not exhibiting a typical MM morphology is because we are likely seeing this SNR expanding into a density medium with a large density gradient of an angle of $\sim$45$^{\circ}$ \citep{lazendic2005}. The progenitor ($\sim$35$-$40 M$_{\sun}$) of G349.7$+$0.2 possibly exploded within the ring-like CO shell having an average density of 35 cm$^{-3}$, where the location of the explosion may be between a dense MC clump and the boundary separating the CO shell from the ISM. The radio continuum morphology shows two rings, where the smaller ring representing the part of the shell expanding into the higher density medium of the MC clump. While this part of the SNR shell might be expanding in the direction away from the observer, it is interacting with the MC clump producing 1720 MHz OH maser emission. The side of the SNR shell expanding towards the observer, on the other hand, is moving into a less dense region inside the CO shell and a smaller section of it might break out of the CO shell into the ISM. 

The stellar wind of the progenitor forms a CIE plasma in the dense circumstellar medium (CSM). When the blast wave breaks out of the dense CSM and expands adiabatically into a cavity or into the rarefied ISM, it causes rapid cooling and the formation of the RP \citep{itohmasai1989, moriya2012, shimizu2012}. So, we would expect the electron cooling for G349.7$+$0.2 to happen through the adiabatic expansion mechanism, when a part of the shell closer to the observer has expanded into a lower density CO shell and then broke out into the rarefied ISM.
We searched for RRC in the {\it Suzaku} data from the E and W regions of G349.7$+$0.2. However, we did not detect any RP features in their spectra.
The total column density calculated from the atomic and molecular gas data was obtained from a circular region covering the X-ray remnant  and it was found as N(total) = 7.1 $\times$ 10$^{22}$ cm$^{-2}$, where $N_{\rm H}$ value of the E and W regions (Table \ref{table_1}) were found to be higher than the N(total) value. The column density value calculated for the E region of G349.7$+$0.2 was found to be lower than the $N_{\rm H}$ value for the W region, while the W region represents the portion of the SNR shell expanding inside the CO shell in other directions. The {\it Suzaku} analysis revealed the enhanced ejecta emission of Si, Ca, and Fe in the E region pointing to the reverse shock produced by the SNR-MC interactions.

To understand whether the dominating gamma-ray emission model for G349.7$+$0.2 is leptonic or hadronic, we modeled all of the emission processes at GeV energies individually. We have adjusted the model parameters to explain the observed fluxes such that we get a set of parameters for each different dominant process. However, some of the parameters estimated were either non-physical or inconsistent with the observed values. Based on these parameters, we could get some insight into the emission mechanisms inside G349.7$+$0.2.

For the case of bremsstrahlung dominated model, the estimated electron to proton ratio was found to be much higher than the value ($\sim$0.01) observed on Earth. In the case of IC process, we found that the number of ambient proton density has to be less than or equal to 0.1, which is inconsistent with the measured proton density values reported in this paper. Moreover, electron to proton ratio was found to be much higher than the observed value. Therefore, both of these leptonic emission processes are insufficient to explain the observed gamma-ray fluxes.

In the case of hadronic model, we considered observed electron to proton ratio to be about 0.01, and found that the hadronic model could explain the observed fluxes well without having any inconsistency in the model parameters. The magnetic field was found to be 150 $\mu$G, which is close to the magnetic field value of one of the OH masers measured by \citet{brogan2000}.

The total gamma-ray luminosity was found as L = 1.58 $\times$ 10$^{36}$ erg s$^{-1}$, similar to the first GeV-emitting SNRs that were discovered by {\it Fermi}-LAT, e.g. IC443 \citep{abdoIC4432010}, W51C \citep{abdoW51C2009}, W44 \citep{abdoW442010}, W49B \citep{abdoW49B2010}, and 3C 391\citep{ergin2014}, all of which are MC interacting MM SNRs with gamma-ray luminosities higher than 10$^{35}$ erg s$^{-1}$.

In order to explain the reported GeV-TeV gamma-ray flux of G349.7$+$0.2, \citet{abramowski2014} estimated that the total proton energy is unreasonably large for an ambient proton density of 1.7 cm$^{-3}$. As a result, they concluded that the TeV gamma-ray emission possibly cannot arise from the whole SNR shell, but from a region where the SNR is interacting with an MC. This conclusion also supports our work, which we have done in this paper considering a SNR-MC interaction. They also reported that a leptonic model is strongly disfavored when trying to explain the observed gamma-ray spectrum. This results also agrees with our conclusion in this paper.

One possible scenario for explaining the observed gamma rays is that G349.7$+$0.2 has evolved inside an MC having dense clumps. The SNR has primarily evolved inside the inter-clump medium, compressing and shocking the dense material throughout its passage. During this process, it heats the material to X-ray temperatures, as well as accelerates the particles through diffuse shock acceleration. The particles can be either reaccelerated cosmic rays trapped in the MC and the shell of the SNR or freshly accelerated protons entering the radiatively compressed MC. In this scenario, since the crushed clouds are thin, multi-GeV particles can escape from the shocked MC, which might be the reason of seeing a break in the proton spectrum \citep{blandfordcowie1982, uchiyama2010}. In an alternative scenario, escaping relativistic protons reach a nearby unshocked MC and produce $\pi^{\circ}\!$-decay gamma rays. For this scenario to happen, there must be GeV/TeV sources found outside the radio shell of G349.7$+$0.2 that could produce these escaping protons \citep{gabici2009}. However, there are no nearby cosmic ray sources to G349.7$+$0.2.

All of the interacting SNRs having RP plasma, where most of them also presenting hadronic gamma-ray emission, are of MM-type. Finding RRC features and gamma rays dominated by the hadronic emission would strongly suggest that G349.7$+$0.2 is an MM type SNR. We found that G349.7$+$0.2 has a gamma-ray spectrum, which is likely hadronic in nature, but lacks RP plasma.

\section{Conclusion}
G349.7$+$0.2 is expanding in a dense medium of MCs and emitting gamma rays by interacting with clumps of molecular material. We analyzed GeV gamma rays from G349.7$+$0.2 and found that the observed gamma-ray spectrum is likely hadronic in origin and is following the spectrum of parent protons, a BPL distribution with spectral index parameters of $\alpha$ = 1.89 and $\beta$ = 2.42 and a spectral break at $\sim$12 GeV. This suggests that protons are accelerated to high energies, possibly at the region of SNR shell interacting with the dense MC clump. Although the best-fit location of the gamma-ray emission is slightly offset from the center of the X-ray remnant, it is closer to the E region. Both the locations of the five OH masers and the whole E region of the X-ray remnant are within the 10$\sigma$ contours of the GeV gamma-ray TS distribution. The low breaking energy of the proton spectrum can be explained either by higher energy protons being already escaped through the crushed MC or the SNR shell at the earlier evolutionary stages of the SNR, or by the SNR shock expanding in a dense medium being slowed down by this medium.
We searched for RP in G349.7$+$0.2, whose MM-nature is still unclear, by studying the plasma structure of two of its regions using the {\it Suzaku} data. We found that the plasma in the E region is in ionization equilibrium, while the plasma in the W region is in the NEI state, showing that none of these regions are in the RP phase.

\acknowledgments
We are grateful to J. G. Magnum for sharing the CO data cubes. We thank A. Bay\5rl\5 for sharing his {\it Fermi}-LAT analysis scripts. T. Ergin acknowledges support from the Scientific and Technological Research Council of Turkey (T\?B\3TAK) through the B\3DEB-2232 fellowship program. A. Sezer is supported by T\"{U}B\.{I}TAK through the B\.{I}DEB-2219 fellowship program. F. G\7k acknowledges support by the Akdeniz University Scientific Research Project Management. E. N. Ercan thanks to Bogazici University for the financial support through the BAP project-5052.
$~$

$~$

Facilities: \facility{Fermi}, \facility{Suzaku}, \facility{NRAO}.

\vspace{0.8cm}



\end{document}